\documentclass[twocolumn,superscriptaddress,preprintnumbers,amsmath,10pt,aps,prl,nobibnotes,longbibliography]{revtex4-1} 
\usepackage{amsmath}
\usepackage[ruled,vlined]{algorithm2e}
\usepackage{amsfonts}
\usepackage{bm}
\usepackage{xcolor}
\usepackage{verbatim}
\usepackage{graphicx}
\usepackage{siunitx}
\usepackage{upgreek}
\usepackage{lipsum}

\begin{document}

\title{Shrinking a gradient index lens antenna system with a spaceplate}

\author{Michal~Mrnka}
\email{M.Mrnka@exeter.ac.uk}
\affiliation{Department of Physics and Astronomy, University of Exeter, Exeter, EX4 4QL, UK.}
\author{Thomas~Whittaker}
\affiliation{Wolfson School of School of Mechanical, Electrical and Manufacturing Engineering, Loughborough University, Leicestershire, LE11 3TU, UK.}
\author{David~B.~Phillips}
\affiliation{Department of Physics and Astronomy, University of Exeter, Exeter, EX4 4QL, UK.}
\author{Euan~Hendry}
\affiliation{Department of Physics and Astronomy, University of Exeter, Exeter, EX4 4QL, UK.}
\author{Will~Whittow}
\affiliation{Wolfson School of School of Mechanical, Electrical and Manufacturing Engineering, Loughborough University, Leicestershire, LE11 3TU, UK.}


\begin{abstract}

\noindent{The miniaturisation of optical systems is an ongoing challenge across the electromagnetic spectrum. While the thickness of optical elements themselves can be reduced using advances in metamaterials, it is the voids between these elements -- which are necessary parts of an optical system -- that occupy most of the volume. Recently, a novel optical element coined a `spaceplate' was proposed, that replaces a region of free space with a thinner optical element that emulates the free-space optical response function -- thus having the potential to substantially shrink the volume of optical systems. While there have been a few proof-of-principle demonstrations of spaceplates, they have not yet been deployed in a real-world optical system. In this work, we use a bespoke-designed spaceplate to reduce the length of a gradient index (GRIN) lens microwave antenna. Our antenna is designed to operate at 23.5~GHz, and the incorporation of a nonlocal metamaterial spaceplate enables the distance between the antenna feed and the GRIN lens to be reduced by almost a factor of two. We find the radiation patterns from a conventional and space-squeezed antenna are very similar, with a very low cross-polarisation, and only a minor increase in the side-lobe levels when introducing the spaceplate. Our work represents a first example of a spaceplate integrated into a functional optical system, highlighting the potential for this concept to reduce the physical size of optical systems in real-world applications.}

\end{abstract}

\maketitle

\noindent {\bf \large 1. Introduction}\\ 

\begin{figure}[ht!]
\centering
\includegraphics[width=1\linewidth]{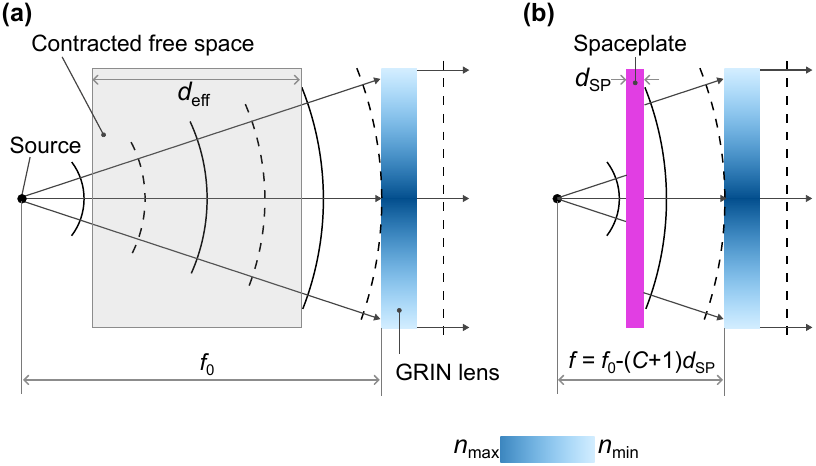}
\caption{\textbf{Principle of a spaceplate.} \textbf{(a)} A GRIN lens illuminated by a point source feed located at the focus. \textbf{(b)} The focus distance is reduced compared to the focal length when a spaceplate is introduced between the lens and the source.  }
\label{fig:1}
\end{figure}

\noindent{In an optical system, the optimum distance between the elements is directly embedded in the function the elements perform. Taking a simple example: if a lens is designed to collimate a beam from a point-like source (e.g. the feed of an antenna), the distance between the source and the lens must be exactly one focal length -- see Fig.~\ref{fig:1}(a). In such a system, the trivial solution to reduce the distance between source and lens is by using a stronger lens of shorter focal length (assuming one can modify the source to match the numerical aperture, NA, of the lens). However, as the focal length is reduced, design trade-offs arise, encompassing higher levels of aberration in imaging systems, and impaired polarization and side-lobe level performance for collimation optics in antenna systems. 

}

Shrinking the distance between optical elements, while preserving their focal lengths, is a challenging problem. The reduction of the axial distances in optical systems can be achieved by folding a beam via polarization dependent reflections, which is the method used in pancake optics in the optical frequency region~\cite{LaRussa1978,Chang2020, Cakmakci2021} , and folded reflect/transmit-arrays  in the radio part of the spectrum~\cite{Pilz1998, Anderton2006, Ge2018, Zhu2022, Lei2022, Bian2023, Wang2024}. These solutions are optimized for certain input polarization and rely on converting polarization states, therefore they do not support operation with arbitrary polarization.
Moreover, these solutions cannot be employed to reduce the dimension of existing systems, as they must be designed together with the focusing elements as a single system, which leads to an increase in complexity of the design process.

Recently, it has been proposed that the distances between optical elements can also be contracted independently of the optical elements themselves through novel approaches in the emerging field of nonlocal metamaterials \cite{Reshef2021, Monticone2022, Miller2023, Mrnka2022, Shao2024}. Here, nonlocality refers to the operating principle of a optical system acting on individual Fourier components of an incident field, i.e. the plane-wave components propagating at given angles in free space, rather than on field components at specific positions. By designing the angular response of these nonlocal materials to match the propagation of radiation in free space, one can design `spaceplates' to mimic free-space diffraction, as first proposed in \cite{Reshef2021, Guo2020}. Spaceplates can replace a slab of free space with a much thinner metamaterial, effectively shrinking the thickness of optical systems -- see Fig.~\ref{fig:1}(b). Due to their operation in momentum space instead of real space, spaceplates are complementary components to focusing elements such as lenses, curved mirrors and by extension reflect/transmit arrays. However, one should note that the behaviour presented in Fig.~\ref{fig:1} 
occurs without any changes to the properties of propagating beams nor any modifications to the focusing power of elements in the original system. Thus, the design of a spaceplate can be completely separated from the design of the rest of the system, unlike in the folded systems mentioned above. 


The basic metric of a spaceplate, which describes its space shrinking power, is the \textit{compression factor C}. This can be defined as the thickness of free space the spaceplate mimics, $d_{\mathrm{eff}}$, to the thickness of the spaceplate itself, $d_{\mathrm{SP}}$:

\begin{equation}
    C =  d_{\mathrm{eff}}  / d_{\mathrm{SP}}.
\end{equation}

\noindent There are fundamental bounds that tie the compression factor together in a trade-off with numerical aperture and  frequency bandwidth \cite{Mrnka2022, Shastri2022}.  In practice, limits on the compression factor are imposed by the required numerical aperture, frequency bandwidth and the physical limits on the lateral size of the elements, i.e. the larger the compression, the smaller the operational bandwidth. If larger bandwidth is required, two recent works \cite{Mrnka2023, Pahlevaninezhad2023} have shown that by dividing a wide bandwidth into a two or three isolated bands allows much higher compression within the sub-bands compared to a structure that continually covers the full band.

In this paper, we introduce the first experimental demonstration of a nonlocal spaceplate contracting the size of a real-world optical system 
-- a microwave gradient index (GRIN) lens antenna \cite{Zhang2016, Zhang2022, Xu2022}. 
The spaceplate is formed by three suitably coupled Fabry-Pérot cavities employing partially reflecting metasurfaces. The spaceplate with thickness of 17.1~mm acts as 120~mm of free space, giving a compression factor $\sim7$. The spaceplate is inserted between the lens and the feed antenna, bringing the focus closer to the lens from 200~mm to 117~mm, while keeping the focal length of the system unchanged. The distance between the lens and the feed antenna is as a result reduced from 150 to 67~mm. The final device is characterized by means typical in antenna metrology - the far-field radiation patterns, cross-polarization, directivity, gain and efficiency are measured in a near-field antenna range and the optical systems with and without the spaceplate compared. We find that the directivity of the space-compressed antenna is comparable to the uncompressed one, while  the efficiency and thus the gain of the compressed antenna is reduced by $\sim$35\%. Finally, we discuss possible routes to improvement in the future.  \\

%
\begin{figure*}[ht!]
\centering
\includegraphics[width=0.8\linewidth]{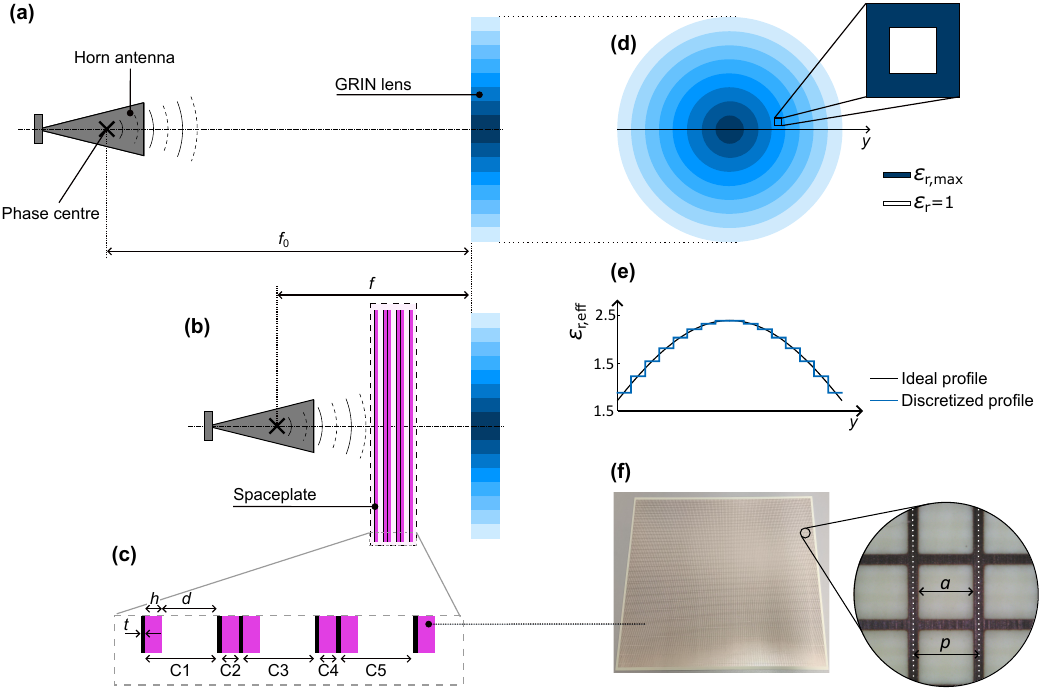}
\caption{\textbf{Principle and experimental setup.} \textbf{(a)} A GRIN lens illuminated by a horn antenna with its phase centre located at the focus of the lens. \textbf{(b)} The focus distance differs from the focal length of the GRIN lens when the spaceplate is introduced. \textbf{(c)} Detail of the multilayer spaceplate. \textbf{(d)} Structure of the GRIN lens. \textbf{(e)} Discretized permittivity profile of the GRIN lens. \textbf{(f)} Photograph of a single mirror of the spaceplate with detail of the periodic metallic mesh structure.}
\label{fig:2}
\end{figure*}

\vspace{10pt}
\noindent {\bf \large {2. GRIN lens antenna}}\\
A GRIN lens antenna is designed to demonstrate the space-squeezing capability of the spaceplate. The antenna is composed of a dielectric GRIN lens and a rectangular, linearly-polarized, horn-antenna source (Flann standard gain horn 20240) with its phase center co-located with the lens's focus. The phase center represents a virtual point associated with the horn antenna which appears to be the point-like source of the spherical wavefronts emanating from the horn antenna.  The horn has aperture dimensions of 52 and 38~mm and produces a pencil beam radiation pattern with a measured 3dB beamwidth of 8.5\,$^\circ$ and 8.2\,$^\circ$ in E and H planes, respectively with a measured directivity of 26.1~dBi at 23.5~GHz. The axial distance between the aperture of the source and the lens is 150~mm. The GRIN lens is designed with a focal length of $f_0$ = 200~mm, as the phase centre of the horn used for measurements is 50~mm further from the source aperture--see Fig.~\ref{fig:2}(a). The lens is 100~mm in diameter with 20~mm thickness and its edges are illuminated at -17.1~dB and -14.8~dB below the axial illumination level in the E and H planes, respectively. The numerical aperture of the lens is $\sim$0.24 ($\theta_\mathrm{max}$ = 14\,$^\circ$). 

The lens is fabricated from high impact polystyrene (HIPS) by a Raise 3D Pro 2 3D printer. A sample tile of HIPS manufactured by the same 3D printer is characterized in a Fabry–Pérot open resonator (see Supplementary information, Fig.~\ref{fig:SupplementaryMeasureDips}); at 23~GHz the sample had a characterized relative permittivity value of 2.45 and a dielectric loss tangent of $1.1\times10^{-3}$. The gradient index profile is discretized into 8 concentric rings each with its own relative permittivity value. From the centre most ring to the outer most, the relative permittivity value of each ring decreases, as described in \cite{Zhang2016}, see Fig.~\ref{fig:2}(f). The effective permittivity is  controlled by varying the volume percentage of air-to-plastic for each ring--see Fig.~\ref{fig:2}(e); this is achieved through the infill percentage parameter in the 3D printer's slicing software. The volume fraction can be calculated using $v={(\varepsilon_\mathrm{r, eff}-1)}/{(\varepsilon_\mathrm{r, max}-1)}$; where $\varepsilon_\mathrm{r, eff}$  and $\varepsilon_\mathrm{r,max}$ are the desired relative permittivity value and the bulk relative permittivity of the material, respectively \cite{Zhang2016}. The unit cell size is controlled by the 3D printer's slicing software but is largely determined by the 3D printer's extrusion nozzle diameter,  the volume fraction and the infill pattern; for the GRIN lens a 0.4~mm nozzle diameter is used with a 'grid' infill pattern. Fig.~\ref{fig:SupplementaryLensPhoto} of the supplementary information shows the photograph of the printed lens with a detail of the relative permittivity profile and with the detail of the porous structure. \\


\noindent {\bf \large 3. Spaceplate design}\\
\label{sec:spaceplate}
A practical way to design microwave spaceplates is based on Fabry-Pérot cavities \cite{chen2021, Mrnka2022}. These can be stacked under suitable coupling conditions \cite{chen2021, Mrnka2023} effectively extending the amount of free space that can be replaced by the device. 
It is the dispersion of the Fabry-Pérot cavity close to its resonance that shows a similar behaviour as free space within limited numerical aperture, determined by Q-factor of the cavity. Under a small angle approximation, both, the transmitted phase of the FP cavity and of free space follow a quadratic dependency on incident angle.

The structure of the proposed spaceplate is given in Fig.~\ref{fig:2}(c) and (f). Semi-transparent mirrors with specifically designed reflectances based on metallic hole-arrays deposited on dielectric substrates are used to form the microwave cavities of the spaceplate. The three space-squeezing cavities C1, C3 and C5, partially filled with air and partially by the dielectric, have electrical thickness of $\approx\lambda/2$ and thus operate near the constructive interference condition. On the other hand, the dielectric-filled cavities C2 and C4 are electrically shorter and are used to set appropriate coupling level between the air-filled cavities \cite{chen2021}.

\begin{figure*}[ht!]
\centering
\includegraphics[width=1\linewidth]{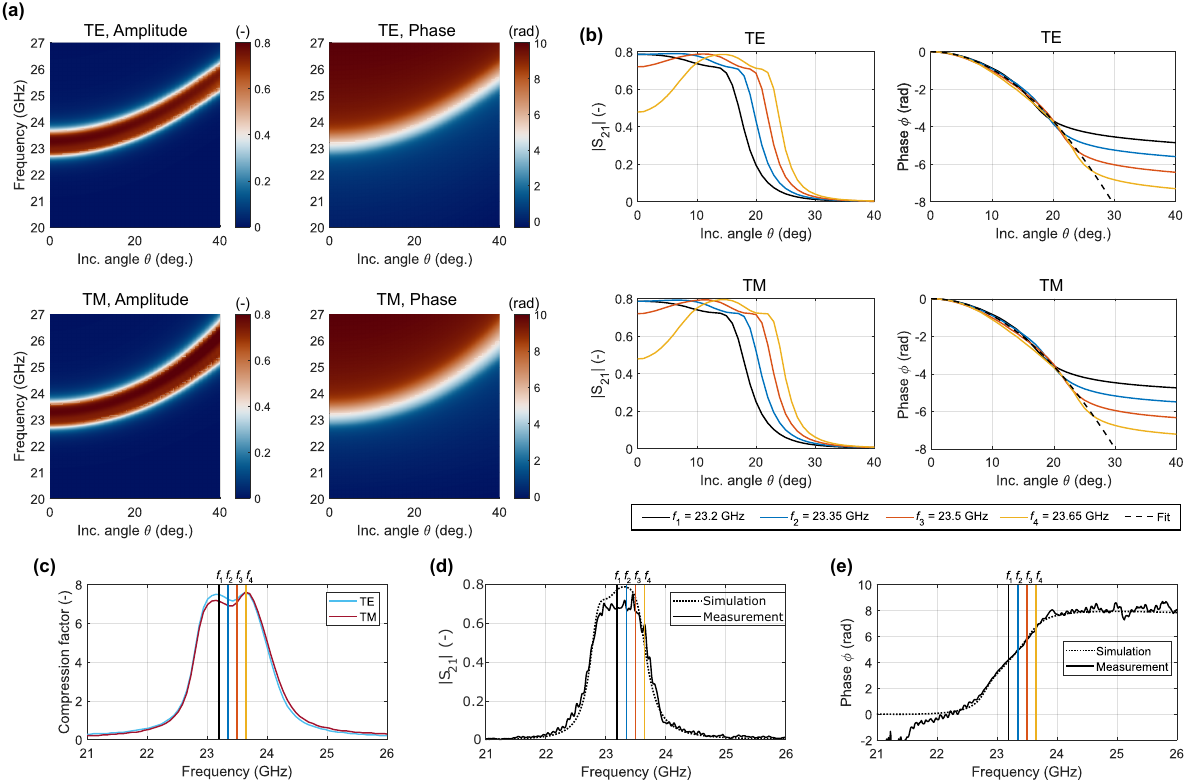}
\caption{\textbf{Spaceplate performance}. \textbf{(a)} Simulated amplitude and phase of the transmission coefficient as a function of angle and frequency for the two polarization states. \textbf{(b)} Simulated amplitude and phase of the transmission coefficient as a function of angle at four selected frequency points. \textbf{(c)} Simulated compression factor $C$ calculated from the dispersion plots in (a).  Measurement of the transmission coefficient at normal incidence - amplitude \textbf{(d)} and phase \textbf{(e)} and comparison with simulation results.}
\label{fig:Simulations}
\end{figure*}

The individual mirrors are fabricated through a standard printed circuit board (PCB) process from Rogers Ro4350B microwave substrate. The relative permittivity of the dielectrics is 3.66 and the loss tangent 0.0037 as defined by manufacturer at 10~GHz frequency \cite{RO4350B}. The thicknesses of the dielectric layers and the copper cladding are $h=$1.524~mm and $t$ = 0.035~mm, respectively. We design the parameters of the single sheets through unit cell modelling in commercial finite-element software Ansys HFSS. Fig.~\ref{fig:SupplementarySingleSheet} in supplementary document presents the simulated reflectance as  functions of frequency at normal incidence for square holes with size $a = 1.26$\,mm and period $p = 1.5$\,mm. At 23.5~GHz, the transmission of the single layer is $\sim$0.19 and reflection $\sim$0.81. Supplementary information section \S 3 also outlines the change of reflectance with incidence angle. A single Fabry-Pérot cavity, which represents the building block for our multi-cavity design,  is formed by two such mirrors and its performance is also described in Supplementary information section \S 3. The compression factor of the single cavity is evaluated from the finite element simulations as $C = 6.9$ compared to the theoretical value $C = -\pi/(2\,\mathrm{ln}R) = 7.45$, which does not account for the losses and changes in reflectance with incident angle.

The spaceplate explored in this paper is assembled by three single-cavity FP resonators stacked back-to-back. Each of the three resonant cavities (i.e. C1, C3, C5) is formed in the region between two successive metallic mirrors - in a space partially filled with the dielectric substrate and partially with air. Considering the  reflection phase of the meshes, both the electrical and actual thickness of the cavities is $<\lambda/2$. The coupling cavities C2 and C4 are filled by the dielectric substrate and for convenience in the assembly process their dimensions are not further optimized. The thicknesses of the air gaps $d$ are tuned simultaneously to achieve the transmission maximum at about 23.5~GHz for $d$ = 2.6~mm. The total thickness of the spaceplate is thus $d_\mathrm{SP} = 17.15$\,mm.  As shown in Fig.~\ref{fig:Simulations}c, finite element modelling indicates that the compression factor for the three-cavity spaceplate is slightly higher ($C \sim 7.2$) compared to a single isolated Fabry-Pérot spaceplate ($C \sim 6.9$), as a result of the coupling among cavities. This corresponds to a more than a threefold increase of $d_{\mathrm{eff}}$, when the three FP resonators are stacked.

In Fig.~\ref{fig:Simulations}, simulation results of the designed spaceplate are summarized for the two principal polarization states with respect to the surface of the spaceplate - transverse electric (TE) and transverse magnetic (TM). Fig.~\ref{fig:Simulations}(a) shows the dispersion plots, where the amplitude and phase of transmission coefficient are plotted as functions of frequency and angle, showing the approximately parabolic shift of the resonance with angle. 1D angular cuts are given in Fig.~\ref{fig:Simulations}(b) showing the non-flat amplitude across the NA of the spaceplate at four frequency points with a significant loss in transmission at normal incidence when operating further away from the resonance (23.65\,GHz). On the other hand, the phase follows the phase of free space even at this frequency.  Simulated compression factor $C$ as a function of frequency is plotted in Fig.~\ref{fig:Simulations}(c) with the four frequency points highlighted on the plots.

We experimentally validate the properties of the isolated spaceplate by measuring amplitude and phase of its transmission coefficient ($\mathrm{S}_\mathrm{21}$) at normal incidence in a two-lens 4f quasi-optical setup (see Supplementary information, Fig.~\ref{fig:SupplementaryMeasureDips}). The setup consists of two identical standard gain horn antennas from Flann microwave (model no: 20240) each with a mid-band gain of 20~dBi and two identical lenses. The lenses were 3D printed from  HIPS and were designed to have a focal length of 130~mm with a radius of curvature of 75~mm. The spaceplate is introduced mid-way between the lenses. Transmission coefficient with and without the spaceplate is measured and a normalized transmission coefficient is obtained:
\begin{equation}
    \mathrm{S}_{21}^{\mathrm{norm}} = \frac{\mathrm{S}_{21, \mathrm{Sample}}}{\mathrm{S}_{21, \mathrm{Ref}}}.
\end{equation}
where $\mathrm{S}_{21, \mathrm{Sample}}$ represents the measured complex transmission coefficient with spaceplate sample in between the horns and  $\mathrm{S}_{21, \mathrm{Ref}}$ the reference transmission coefficient with the sample taken out. As a result of this quasi-optical calibration, the measured normalized phase is referenced to a slab of air (free space) with the same thickness as the spaceplate. To remove this effect and to have the phase normalized to reference path length of 0\,mm, we have to introduce the multiplicative factor $\mathrm{e}^{\mathrm{i} k\,d_\mathrm{SP}}$ and the final normalized transmission coefficient is given as:
\begin{equation}
    \mathrm{S}_{21} (f) = \mathrm{e}^{\mathrm{i} k\,d_\mathrm{SP}} \cdot \mathrm{S}_{21}^{\mathrm{norm}} (f).
\end{equation}
Fig.~\ref{fig:Simulations}(d) depicts the measured amplitude of the transmission coefficient  of the spaceplate $|\mathrm{S}_\mathrm{21}|$ as a function of frequency and Fig.~\ref{fig:Simulations}(e) shows the measured unwrapped phase of the transmission coefficient. Simulation data is included in these plots for comparison. We note that the presented measured results are only approximate since the spaceplate introduces a systematic error into the measurement as described in \cite{MrnkaEsa2022} which is related to the apparent change in the distances in the setup with and without the spaceplate.

To incorporate the spaceplate into the antenna design, the individual mirrors are compressed together using threaded PTFE bolts in the corners of the sheets and the distances between the sheets are set by PTFE spacers with thickness of 2.7$\pm$0.1\,mm. The spaceplate is then introduced mid-way between the horn source antenna and the GRIN lens and the distances between the horn, spaceplate and the lens are reduced to respect the space-compression effect -- see Fig.~\ref{fig:Simulations}(c), thus the distance between the horn and the lens' centre is reduced from 150 to 67~mm. The new focal distance $f$~=~117~mm -- see Fig.~\ref{fig:1}(b) compared to the focal length of the GRIN lens $f_0$ = 200~mm.\\

\begin{figure}[ht!]
\centering
\includegraphics[width=1\linewidth]{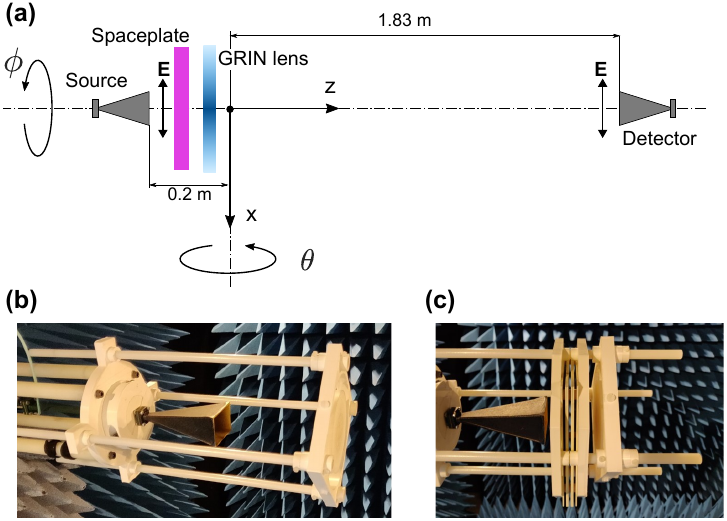}
\caption{\textbf{Experimental setup and realization of the GRIN spaceplate antenna.} \textbf{(a)} Experimental setup for measuring co-polarization radiation pattern.  \textbf{(b)} Photograph of the GRIN lens illuminated by a rectangular horn.  \textbf{(c)} With a spaceplate inserted between the horn and the lens.}
\label{fig:ExperimentSetup}
\end{figure}

\noindent {\bf \large 4. Experiment and results}\\
We evaluate the performance of the space-compressed GRIN antenna by measuring its far-field radiation patterns (directivity) and comparing them to those obtained for the nominal lens antenna without the spaceplate. For completeness we also include the characteristics of the source horn antenna without the lens. The simplified schematic of the experimental setup is given in Fig.~\ref{fig:ExperimentSetup}. The antenna under test -- AUT (horn; horn + GRIN lens; or horn + spaceplate + GRIN lens) is mounted on a rotation stage allowing its rotation in both phi ($\phi$) and theta ($\theta$) axes. A stationary standard gain probe antenna (RF spin DRH40) is placed  at a distance between 1.83--2.03~m away from the AUT, depending on the AUT configuration -- 2.03~m for horn and 1.86~m for horn+lens and for horn+lens+spaceplate. The near-field of the AUT is then mapped on a sphere by measuring the transmission coefficient between the antennas as a function of angles $\theta$ and $\phi$. The transmission coefficient is characterized for two orthogonal polarizations of the probe antenna and the results corresponding to the antenna's co- and cross-polarization radiation patterns are obtained upon near-field to far-field transformation. During the co-polarization measurement, the electric field vector radiated by the source and the polarization sensitivity of the detector are parallel whereas in the cross-polar measurement they are perpendicular to each other.  

\begin{figure*}[ht!]
\centering
\includegraphics[width=0.9\linewidth]{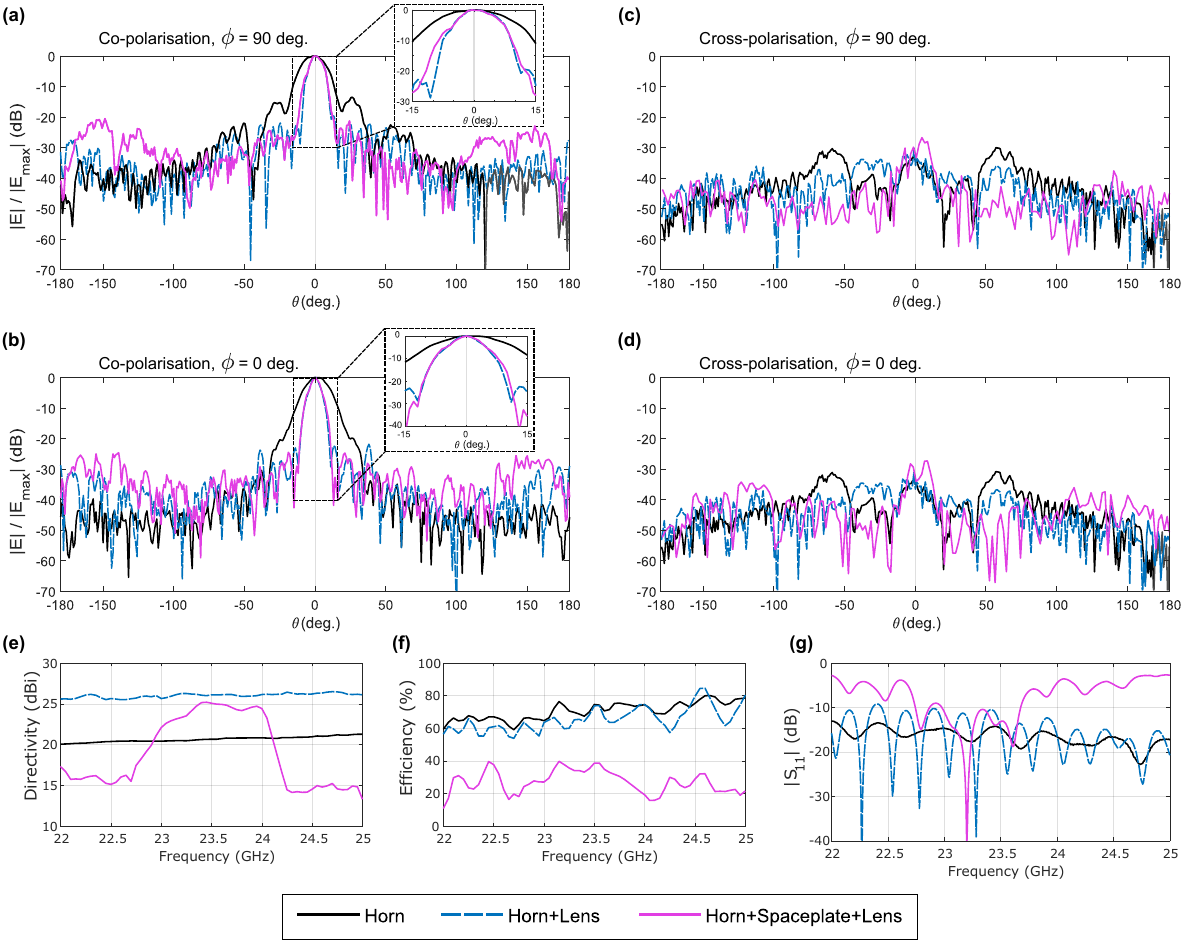}
\caption{\textbf{Experimental results.} Co-polarized component of the radiated field as a function of the rotation angle in the two principal planes of the antenna - E-plane \textbf{(a)}  and H-plane \textbf{(b)} for the three antenna configurations: horn alone, horn + GRIN lens, horn + spaceplate + GRIN lens. Cross-polarized field components on the E- and H- principal planes in \textbf{(c)} and \textbf{(d)}, respectively. Directivity \textbf{(e)}, efficiency \textbf{(f)} and reflection coefficient \textbf{(g)} as functions of frequency.}
\label{fig:ExperimentResults}
\end{figure*}

Fig.~\ref{fig:ExperimentResults} gives the comparison of the co-polar, normalized radiation patterns of the three above mentioned configurations (horn; horn + GRIN lens; or horn + spaceplate + GRIN lens) at 23.5~GHz for the two orthogonal planes $\phi$~=~0\,$^\circ$ and $\phi$~=~90\,$^\circ$ (azimuth and elevation planes), in Figs.~\ref{fig:ExperimentResults}(a) and (b), respectively. We see how the GRIN lens increases the directivity of the source horn, thus reducing beamwidth, and how the introduction of the spaceplate preserves the shape of the main beam of the lens antenna despite the largely reduced separation between source horn and GRIN lens (from 150 to 67~mm). We also observe the gradual increase in the radiation levels outside of the main beam (i.e. reduced spillover efficiency) from the source antenna alone, to the GRIN lens and finally to the spaceplate-squeezed GRIN lens antenna. In case of the lens, these are caused by the reflection at the air-dielectric interface due to the index mismatch. In case of the spaceplate based antenna, the reflection from the spaceplate at angles outside of its numerical aperture is substantial and approaches 100\% for angles over 30\,$^\circ$ -- see Fig.~\ref{fig:Simulations}(b). The prominent sidelobes at angles $\pm$153$\,^\circ$ in the elevation plane of the spaceplate based antenna are actually reflections of the sidelobes of the source horn antenna located at the $\pm$27$\,^\circ$. This is the reason why the lobes are more subtle in the H-plane, where the source antenna has much lower radiation levels in the given direction. Figs.~\ref{fig:ExperimentResults}(c) and (d) show how the cross-polarization level is influenced by the addition of the spaceplate into the optical setup. We see a subtle depolarization of the fields  due to the presence of the spaceplate, but the overall cross-polarization level at the boresight (on-axis direction $\theta$~=~0$\,^\circ$) is close to -30~dB.

In Fig.~\ref{fig:ExperimentResults}, directivity (e),  efficiency (f) and reflection coefficient at the input port of the horn antenna (g) are compared as functions of frequency. The efficiency of the AUT was calculated using the direct comparison method as the gain profile of the probe is known and the loss in the cables can be measured. 
 The reflection coefficient in Fig.~\ref{fig:ExperimentResults}(g) indicates that the spaceplate is well impedance-matched to the rest of the elements in the frequency range of interest (22.7 -- 23.6~GHz); hence it is not the back-reflection from the spaceplate that reduces the efficiency of the antenna.\\

\noindent{\bf \large 5. Discussion}\\
An important consideration when placing a spaceplate close to an electromagnetic source, a feed horn antenna in our case, is its influence on the properties of the source itself via reactive near-field interaction. In such a case the electromagnetic field distribution within antenna's aperture is influenced by capacitive and/or inductive coupling resulting from the physical proximity of the spaceplate. As a result, an alteration of the antenna's radiation pattern follows -- the source antenna and the spaceplate become a single electromagnetic component and are no longer separable in their function. In the experiment with the spaceplate inserted between the lens and the feed horn, the axial distance from the edge of the horn to the spaceplate is 25~mm, which correspond to electrical length of $\sim$2 $\lambda$. This suggests that the spaceplate might in fact be located within the transition between the reactive and radiative near-field of the horn, as this boundary is not uniquely defined, especially not for aperture antennas such as horns \cite{Lee2011}. We thus conclude that the ripple in the main beam of the radiation pattern of the spaceplate enhanced antenna [insets in Fig.~\ref{fig:ExperimentResults}(a), (b)] might be in part caused by the reactive near-field interaction between the horn antenna and the spaceplate. This small effect, however, does not appear to alter significantly the directivity of the antenna: the small reduction in directivity can also be explained by the varying transmittance of the spaceplate within its numerical aperture [see Fig.~\ref{fig:Simulations}(b)], which manifests as a distortion of the main beam of the antenna.

While the directivity of the space-compressed antenna in the frequency band 23.4~GHz - 23.7~GHz is very close to the nominal GRIN lens antenna, the efficiency and thus the gain is reduced by around 35\%. Our finite element modelling indicates that this is mainly the result of the dielectric losses in the substrates of the cavity mirrors. Dielectric substrates with lower loss tangent would reduce this problem. To further rectify this effect, a fully metallic solution would be preferable. This may be more difficult to achieve, as it would require rigid, self supporting metasurfaces with deviations in surface flatness considerably smaller than the wavelength.\\

\noindent{\bf \large 6. Conclusions}\\
In summary, we have experimentally demonstrated a size reduction of a microwave-optics system -- a GRIN lens antenna by means of a  spaceplate. The spaceplate is capable of compressing a slab of free space seven times as thick. By inserting the spaceplate between the GRIN lens and its source horn antenna, we reduce the length between the lens and its focus by almost a factor of two while keeping the focal length of the optics unchanged. We find that the radiation patterns of the squeezed system are very similar to those of the uncompressed antenna, with a very low cross-polarization, similar directivity and a minor increase in the side-lobe levels. A considerable degradation in efficiency of around 35\% is attributed to losses in the dielectric materials used to support the individual metasurfaces. In future, it may be possible to redress the efficiency using low-loss dielectrics or fully metallic metasurfaces, and if achieved we believe that space compression metamaterials will be an important route towards miniaturized optical devices operating in microwave region. Moreover, even though we demonstrate operation of the antenna with linearly polarized radiation, the symmetry of the structure makes it polarization agnostic, thus operation with circular or any general polarization sense is possible. We believe our work furthers the research on spaceplates and validates their application potential.\\


\noindent {\bf \large \textcolor{black}{Acknowledgments}}\\
The authors acknowledge financial support from the Engineering and Physical Sciences Research Council (EP/S036466/1, EP/W003341/1 and EP/030301/1). D.B.P. acknowledges financial support from the European Research Council (804626), and the Royal Academy of Engineering. For the purpose of open access, the author(s) has applied a ‘Creative Commons Attribution (CC BY) licence to any Author Accepted Manuscript version arising.
\\

\noindent {\bf \large Disclosures}\\
 The authors declare no conflicts of interest.\\


\noindent {\bf \large References}\\

\bibliography{references}

\onecolumngrid
\setcounter{equation}{0}
\setcounter{figure}{0}
\setcounter{page}{0}
\renewcommand{\figurename}{FIG.}
\renewcommand{\thefigure}{S\arabic{figure}}
\renewcommand{\tablename}{TABLE}
\renewcommand{\thetable}{S\arabic{table}}
\vspace{50cm}
\noindent{\centering{\LARGE Supplementary Information}\par}
\vspace{5mm}

\noindent{This supplementary document provides further, in-depth information on the design, simulations and measurements of the spaceplate based GRIN antenna.}\\

\noindent{\large \bf \S 1 Experimental setup for material characterization}\\
The photo of the experimental setup used to characterize the material properties of the samples is given in Fig.~\ref{fig:SupplementaryMeasureDips}. Here, the setup is shown during measuring the spaceplate's dispersion at normal incidence -- the results are given in Fig.~\ref{fig:Simulations}(d), (e) of the main paper. The description of the setup can be found in the main paper -- section 2.

\begin{figure}[ht!]
\centering
\includegraphics[width=0.8\linewidth]{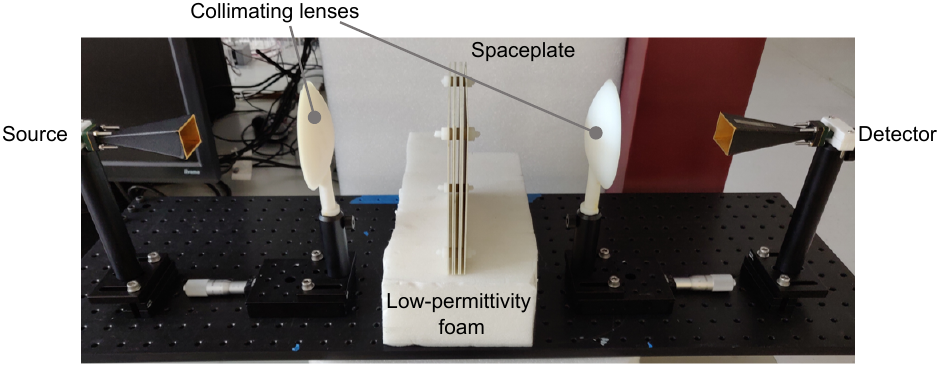}
\caption{\textbf{Experimental setup for measuring frequency dispersion of the spaceplate and the 3D printed materials.} }
\label{fig:SupplementaryMeasureDips}
\end{figure}

\noindent{\large \bf \S 2 GRIN lens design and fabrication}\\
The GRIN lens described in the main paper is composed of eight regions with varying effective permittivity -- the volumetric filling fractions and calculated relative permittivities of the individual regions are given in Tab.~\ref{tab:GRIN_Properties} and the photograph is provided in Fig.~\ref{fig:SupplementaryLensPhoto}

\begin{table}[h!]
    \centering
    \begin{tabular}{|c|c|c|}
    \hline
        Region No.       & Filling fraction    & Calculated $\varepsilon_\mathrm{r}$\\
                        & for 3D printer      & \\ \hline
        1 (inner most)  & 100   & 2.45\\
        2               & 98    & 2.42\\
        3               & 94    & 2.36\\
        4               & 88    & 2.27\\
        5               & 80    & 2.16\\
        6               & 70    & 2.02\\
        7               & 60    & 1.87\\
        8 (outer most)  & 48    & 1.69\\ \hline
    \end{tabular}
    \caption{Calculated properties of GRIN lens regions.}
    \label{tab:GRIN_Properties}
\end{table}

\begin{figure}[ht!]
\centering
\includegraphics[width=0.5\linewidth]{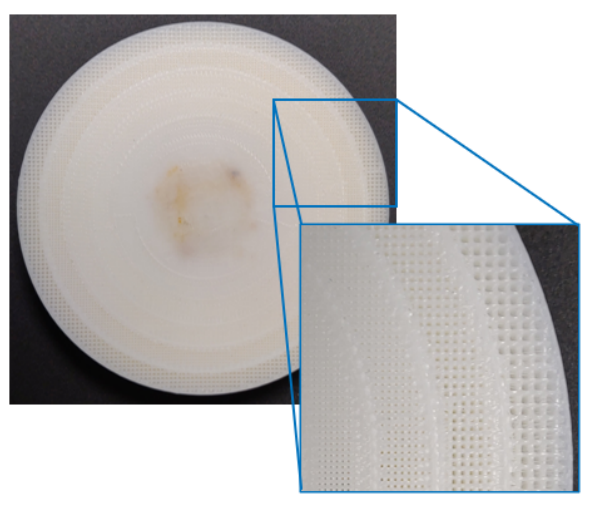}
\caption{\textbf{Photo of the GRIN lens.} }
\label{fig:SupplementaryLensPhoto}
\end{figure}

\noindent{\large \bf \S 3 Spaceplate analysis and design}\\
As the  spaceplate descibed in the main paper is composed of six partially reflecting mirrors and can be decomposed into three individual FP cavities, here we present properties of these building blocks. Fig.~\ref{fig:SupplementarySingleSheet}(a) introduces the unit cell of the hole array structure deposited on microwave substrate with period $p$ and hole size $a$. Fig.~\ref{fig:SupplementarySingleSheet}(b) shows the reflectance of the sheet as a function of frequency for normal incidence and for 30$^\circ$ incident angle. The small difference for TE and TM polarization can be observed arising from the Fresnel reflection coefficient associated with the interface. \\

\begin{figure}[ht!]
\centering
\includegraphics[width=0.75\linewidth]{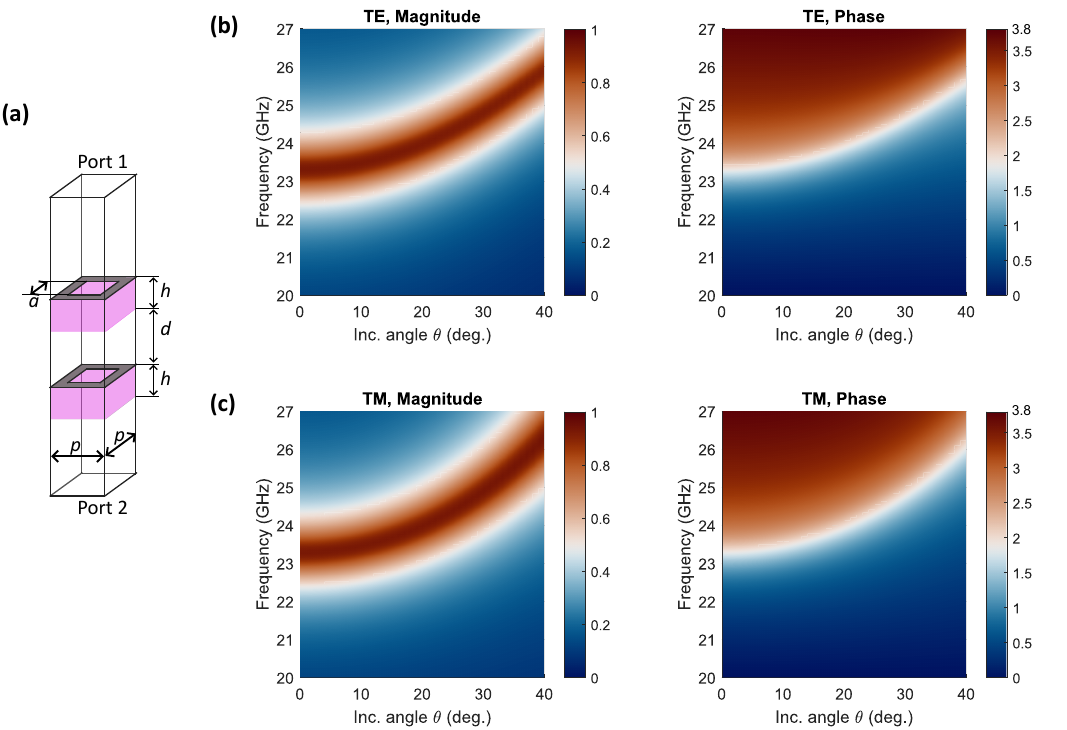}
\caption{\textbf{Simulated properties of a single Fabry-Pérot cavity.}  \textbf{(a)}  Unit cell schematics. Dispersion plots for TE \textbf{(b)} and for TM \textbf{(c)} polarization states showing the magnitude and phase of the simulated transmission coefficient.}
\label{fig:SupplementarySingleCavity}
\end{figure}

By stacking two such partial mirrors, we form a single FP cavity with unit cell given in Fig.~\ref{fig:SupplementarySingleCavity}(a). The thickness of the air gap between the sheets is denoted here as $d$. The magnitude and phase of the simulated transmission coefficient is given in Fig.~\ref{fig:SupplementarySingleCavity}(b) and (c) as functions of incident angle and frequency. The simulations were performed in finite-element solver Ansys HFSS.

\begin{figure}[ht!]
\centering
\includegraphics[width=0.8\linewidth]{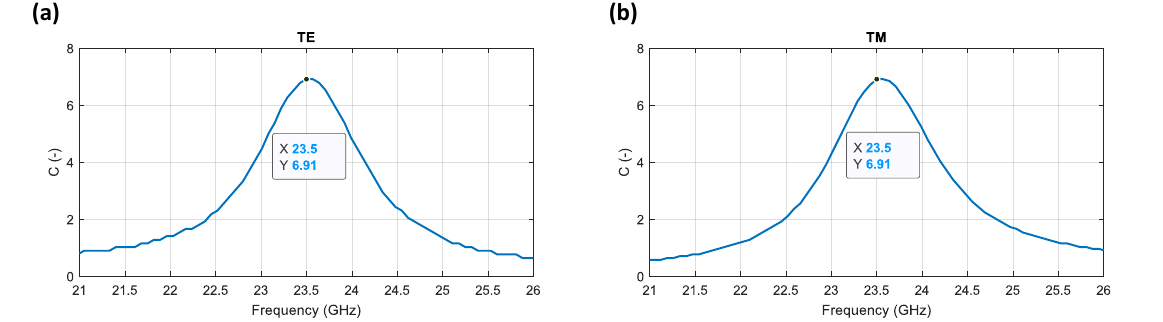}
\caption{\textbf{Calculated compression factor of a single Fabry-Pérot cavity spaceplate.} \textbf{(a)} For TE polarization. \textbf{(b)} For TM polarization.}
\label{fig:SupplementarySingleCavityCompression}
\end{figure}

We evaluate the effective thickness of free space that the spaceplate replaces from fitting a free-space dispersion $\arg\{\exp\left(\mathrm{i}k\,d_{\mathrm{eff}} \cos{\theta}\right)\}$ to the the simulated phase dispersion. This fitting is undertaken for each frequency  over the range of angles corresponding to the operating range of the spaceplate i.e. numerical aperture $\mathrm{NA} = \sin{\theta_\mathrm{max}}$, where $\theta_\mathrm{max}$ represents the maximum incident angle accepted by the spaceplate, in our case $\theta_\mathrm{max} = 20^\circ$. This value is slightly higher than the maximum angle, corresponding to the marginal ray connecting the centre of horn's aperture and the rim of the lens (18.4\,$^\circ$) to improve the transmission modulation across the NA.
The result of the fitting is the distance $d_\mathrm{eff}$ and by knowing the thickness of the spaceplate $d_\mathrm{SP}$ we can evaluate the compression factor $C = d_\mathrm{eff}/d_\mathrm{SP}$. In the case of a single cavity spaceplate from Fig.~\ref{fig:SupplementarySingleCavity}(a), $d_\mathrm{SP}$ = 5.7~mm and the compression factor as a function of frequency is given in Fig.~\ref{fig:SupplementarySingleCavityCompression}(a) and (b) for TE and TM polarization state, respectively.

\begin{figure}[ht!]
\centering
\includegraphics[width=0.6\linewidth]{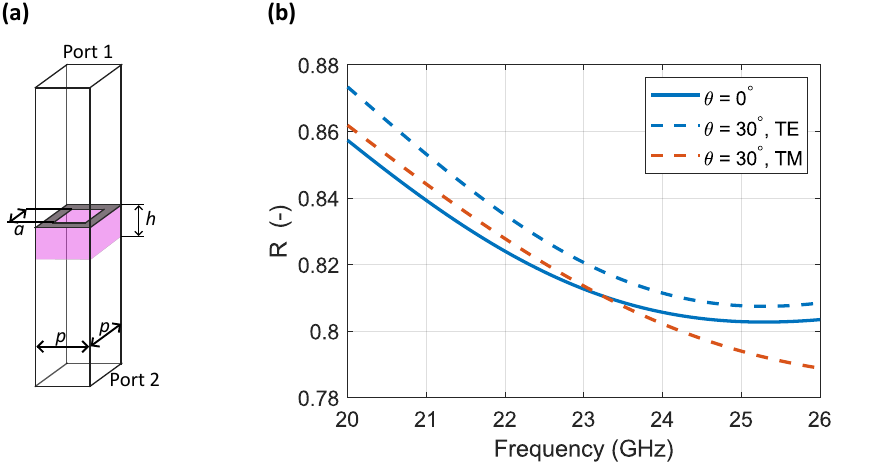}
\caption{\textbf{Simulated properties of individual mirrors.} \textbf{(a)}  Unit cell \textbf{(b)} Reflectance as a function of frequency.}
\label{fig:SupplementarySingleSheet}
\end{figure}

\end{document}